\documentclass[aps,prd,floats,floatfix,amssymb,twocolumn,superscriptaddress,nofootinbib]{revtex4-2}

\usepackage{mathptmx}
\usepackage{physics}
\usepackage{amsmath}
\usepackage{amsfonts,amssymb}
\usepackage{graphicx,graphics}
\usepackage{hyperref}
\usepackage{ulem}
\usepackage{color}
\usepackage{mathrsfs}
\usepackage{bbold}

\hypersetup{
    colorlinks=true,
    citecolor=blue,       
    linkcolor=blue,      
    urlcolor=cyan       
}

\makeatletter
\renewcommand{\tagform@}[1]{(\textcolor{black}{#1})}
\makeatother

\begin{document}

\title{Unstable mode and the Unruh-DeWitt detector}

\author{Bruno S. Felipe}
\email[]{brunfeli@ifi.unicamp.br}
\affiliation{Instituto de F\'isica Gleb Wataghin, Universidade Estadual de Campinas, 13083-859 Campinas, S\~ao Paulo, Brazil}

\author{Jo\~ao P. M. Pitelli}
\email[]{pitelli@unicamp.br}
\affiliation{Departamento de Matem\'atica Aplicada, Universidade Estadual de Campinas, 13083-859 Campinas, S\~ao Paulo, Brazil}%

\begin{abstract}
We investigate the quantization of a single unstable mode in a real scalar field subject to a Robin boundary condition in (1+1)-dimensional half-Minkowski spacetime. The instability arises from an imaginary frequency mode—analogous to that of the inverted harmonic oscillator—requiring the rigged Hilbert space formalism for consistent quantization. Within this framework, the unstable mode is naturally described as a well-defined decaying (or growing) quantum state with a characteristic mean lifetime. We investigate its physical consequences via the response of an Unruh–DeWitt detector along static, inertial, and uniformly accelerated trajectories. For static and inertial observers, the detector response exhibits a Breit–Wigner resonance profile, with a decay width determined by the unstable frequency and a Doppler factor. In the Neumann limit, infrared divergences emerge from arbitrarily low-frequency modes. Interestingly, for accelerated detectors, the response acquires a nontrivial dependence on acceleration, and the Neumann limit yields a finite, oscillatory signal rather than a divergence, suggesting that acceleration can act as an effective infrared regulator.

\end{abstract}

\maketitle

\section{Introduction}

The theory of quantum fields in curved spacetime offers a rigorous framework for studying field quantization in backgrounds governed by general relativity. In this approach, spacetime $(\mathcal{M},g)$ is treated as a classical entity, while the fields, namely $\phi$, are quantized ignoring potential backreaction effects~\cite{birrell,fulling,wald1}. In particular, in stationary spacetimes, the presence of a smooth timelike  Killing vector field allows the decomposition of field solutions into positive- and negative-frequency modes $\{u_i, u_i^{\ast}\}$, enabling quantization through standard commutation relations between the field and its canonical conjugate.

However, certain scenarios challenge this conventional framework. In cases where the field is non-minimally coupled to the spacetime curvature or under specific mass coupling conditions, an effective potential can induce instability in the classical field solution~\cite{lima, landulfo, grain-barrau, lima-matsas-vanzella, lima-vanzella, lasenby}. These instabilities manifest as modes characterized by an imaginary frequency, $\Im{\omega} = \gamma \neq 0$, leading to time evolution dominated by a real exponential that grows unboundedly at large times. As a result, the presence of such modes renders the splitting into positive and negative frequencies meaningless, preventing the established quantization procedure from defining a natural Fock space for particle-content interpretation~\cite{schroer-swieca}.

A related issue arises in the context of static, but non-globally hyperbolic spacetimes, where the absence of a Cauchy surface requires an appropriate boundary condition (BC) at the edge of the spacetime~\cite{wald,ishibashi-wald,helliwell} (for spacetimes with a naked singularity or even a conformal infinity). In such cases, the choice of boundary condition plays a crucial role in defining sensible dynamics for classical fields, as well as the vacuum state that becomes dependent on both the timelike Killing field and the BC. Depending on the imposed boundary condition, the spectrum of the theory may lack a lower bound, giving rise to unstable modes, often referred to as bound states (for examples in $\mathrm{AdS}$, see Refs.~\cite{dappiaggi-ferreira,troost}), and preventing the definition of a ground state with well-defined positive energy at a quantum level.

In a previous work~\cite{felipe-pitelli}, we explored the non-globally hyperbolic half-Minkowski spacetime and showed that a scalar field subject to a Robin boundary condition (RBC) is governed by an effective parabolic potential, leading to the emergence of bound states whose dynamics resemble those of an inverted harmonic oscillator (IHO). This boundary condition was derived by introducing a surface action at the naked singularity, ensuring a well-defined variational principle. Notably, this approach accounts for the energy flow at the singularity and yields a theory with a conserved energy, effectively describing a closed system. These results provide a framework for addressing and quantizing unstable modes, offering insights into their physical interpretation.

Inspired by the behavior of the IHO, in Ref.~\cite{felipe-pitelli} we quantize the unstable mode in an analogous manner. However, unlike the standard harmonic oscillator—commonly used as a basis for canonical quantization—the conventional Hilbert space formulation of quantum mechanics is inadequate to fully capture the mathematical subtleties of the IHO. This system features an eigenvalue problem with either a continuous or purely imaginary spectrum (depending on the chosen basis), whose eigenstates are represented by distributions rather than square-integrable functions. Consequently, a proper quantization of the system and the identification of a suitable Fock representation for the unstable particles require extending the standard framework to the so-called {\it rigged Hilbert space} (RHS) formulation of quantum mechanics.

Building on this foundation, in the present work, we constrain our analysis to a real scalar field defined on the 1+1-dimensional half-Minkowski spacetime. In this setting, we construct a toy model in which a timelike singularity emerges at a certain stage, allowing for energy flux through the singularity. This is effectively equivalent to considering a field initially respecting the Dirichlet boundary condition at $x=0$ and subtly changing to a boundary condition of Robin type after an instant $t=0$ (see the discussion in~\cite{ishibashi-hosoya}). As we will show, this setup gives rise to a single unstable mode, which we quantize within the RHS framework to ensure a mathematically consistent quantum description. To further explore the physical properties of this mode, we employ an Unruh-DeWitt detector to analyze how an observer perceives its effects, thereby gaining insights into the physical implications of the instability and assessing the validity and applicability of the proposed quantization procedure.

The paper is structured as follows. In Sec.~\ref{sec:field-solution} we explore the classical field in (1+1)-half-Minkowski spacetime, and examine how the Robin boundary condition gives rise to a single unstable mode. In Sec.~\ref{sec:quantization}, we revisit the procedure outlined in Ref.~\cite{felipe-pitelli}, emphasizing the key aspects of the RHS framework, the inverted harmonic oscillator, and their role in the quantization of the bound state. In Sec.~\ref{sec:detector}, we apply the Unruh-DeWitt detector model to incorporate the unstable mode and analyze the detector's response for two distinct trajectories. Finally, our conclusions and final observations are presented in Sec.~\ref{sec:final-remarks}. Through this work, we consider natural units $\hslash=c=1$.

\section{Field solution and unstable mode}\label{sec:field-solution}

\begin{figure}[!htb]
    \centering
    \includegraphics[width=0.35\linewidth]{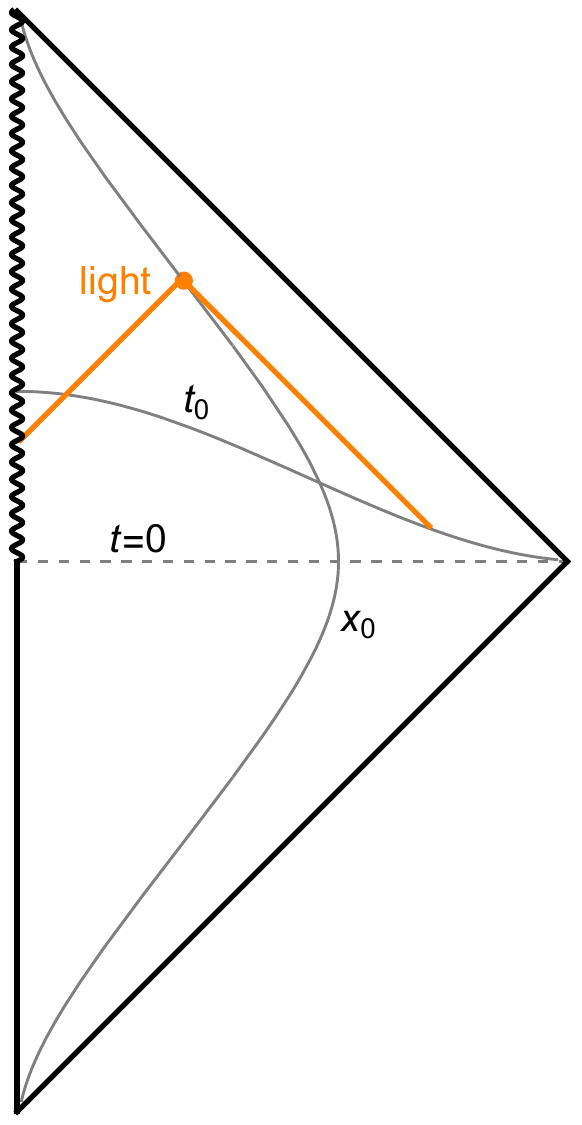}
    \caption{Penrose diagram of the two-dimensional half-Minkowski spacetime. The vertical line in the region $t < 0$ denotes the boundary at $x = 0$, where the Dirichlet boundary condition is imposed. After $t = 0$, the vertical wavy line in the region $t>0$ represents the emergence of a naked singularity, where the Robin boundary condition is imposed.}
    \label{fig:half-mink}
\end{figure}

The two-dimensional half-Minkowski spacetime is characterized by the line element (setting $c=1$)
\begin{equation}
    ds^2=-dt^2+dx^2,
\end{equation}
with $t\in \mathbb{R}$ and $x\in [0,\infty)$. Notably, restricting the spatial coordinate to the half-line renders the spacetime non-globally hyperbolic, thus requiring an appropriate boundary condition at $x = 0$. As illustrated in the Penrose diagram shown in Fig.~\ref{fig:half-mink}, a constant-time surface $t_0$ is no longer a Cauchy surface, since a light signal emitted from (or received at) $x = x_0$ can reach the naked singularity within a finite time interval $\Delta t = \pm x_0$.

For a real massless scalar field $\phi$, an appropriate action coupling the field to a mass parameter $\gamma \geq 0$ at the boundary $x=0$ is given by
\begin{equation}\label{action}
S[\phi]=-\frac{1}{2}\int_{t_1}^{t_2} dt\int_{0}^{\infty} dx  \left(-\dot{\phi}^2+\phi^{\prime\,2}\right)+\frac{\gamma}{2}\int_{t_1}^{t_2} dt  \phi^2,
\end{equation}
where $\dot{\phi}\equiv\partial_t\phi$ and $\phi^{\prime}\equiv\partial_x\phi$. As demonstrated in Ref.~\cite{felipe-pitelli}, the action (\ref{action}) governs the energy flux at $x=0$, ensuring conserved energy for the entire system. Furthermore, by extremizing the action with respect to the field, we find $\delta S = 0$ if and only if
\begin{equation}\label{CC}
    -\ddot{\phi}+\phi^{\prime\prime}=0\quad \text{and}\quad\left.\left( \phi+\frac{1}{\gamma}\phi^{\prime}\right )\right|_{x=0} =0.
\end{equation}
Thus, the equation of motion and the Robin boundary condition naturally emerge from the variational principle applied to the action (\ref{action}). Dirichlet and Neumann boundary conditions are recovered in the limits $\gamma\to\infty$ and $\gamma=0$, respectively.

Eq.~(\ref{CC}) applied to the ansatz $\phi(t,x)=e^{-i\omega t}\psi(x)$ yields
\begin{equation}
    \psi^{\prime\prime}(x) = -\omega^2\psi(x),
\end{equation}
which defines a {\it Sturm-Liouville problem} in $\mathcal{L}^2(\mathbb{R}_{+},dx)$ with the RBC. In this context, the Hermiticity of $\partial_x^2$ only requires $\lambda \equiv \omega^2 \in \mathbb{R}$. For positive eigenvalues, $\lambda > 0$, the basis solutions $\{\psi_1, \psi_2\}$ are given by
\begin{equation}
    \psi_{1}(x)=\frac{\sin(\omega x)}{\omega}\quad\text{and}\quad\psi_2(x)=\cos(\omega x).
\end{equation}
Hence, the general solution in $x$ can be written as a linear combination $\psi = A\psi_1 + B\psi_2$, where the coefficients $A$ and $B$ are related via the boundary condition: $B = -A \omega / \gamma$. 

Before $t=0$, we restrict the field to satisfy the Dirichlet boundary condition, implying $B \to 0$ as $\gamma \to \infty$. Thus, for $t<0$, the normalized mode solutions are given by
\begin{equation}
    u_{\omega}^{(in)}(t,x)=\frac{e^{-i\omega t}\sin(\omega x)}{\sqrt{\pi \omega}}.
\end{equation}
After $t=0$, however, the subtle change to the Robin boundary condition yields the normalized modes
\begin{equation}\label{free-modes}
    u_{\omega}^{(out)}(t,x)=\frac{\gamma e^{-i\omega t}\left(\sin(\omega x)-\frac{\omega }{\gamma}\cos(\omega x)\right)}{\sqrt{\pi\omega}\sqrt{ \gamma^2+\omega^2}}.
\end{equation}

For negative eigenvalues $\lambda < 0$, the Dirichlet boundary condition admits no solution, while the Robin boundary condition is satisfied only if $\omega = \pm i \gamma$, resulting in a point-like energy with the $x$-component adopting a real exponential form. The corresponding mode solution, normalized in $\mathcal{L}^2(\mathbb{R}_+, dx)$, is given by
\begin{equation}\label{imaginary-solution}
    u_{\gamma}(t,x)=\sqrt{2\gamma}\exp(-\gamma x \pm \gamma t).
\end{equation}
At this point, there is no mathematical reason to select time evolution with only $e^{+\gamma t}$ or $e^{-\gamma t}$, and so this state exhibits exponential growth for large $t$, indicating an unstable mode---or equivalently, a bound state---that introduces instability to $\phi$. Furthermore, observe that apart of the normalization, as the boundary parameter approaches the Dirichlet limit ($\gamma \to \infty$), the unstable state vanishes; in the Neumann limit ($\gamma \to 0$), it reduces to a single non-oscillatory zero mode with arbitrarily low energy (see \cite{ishibashi-hosoya}).

The total field can then be written as
\begin{equation}\label{before}
    \phi(t,x)=\int_0^{\infty} d\omega\left(a_{\omega}^{(in)}u_{\omega}^{(in)}(t,x)+a_{\omega}^{(in)\dagger}u_{\omega}^{(in)\ast}(t,x)\right)
\end{equation}
for $t<0$ and
\begin{equation}\label{after}
    \begin{aligned}
        \phi(t,x)&=\int_0^{\infty} d\omega\left(a_{\omega}^{(out)}u_{\omega}^{(out)}(t,x)+a_{\omega}^{(out)\dagger}u_{\omega}^{(out)\ast}(t,x)\right)\\&\quad+u_{\gamma}(t,x)
    \end{aligned}
\end{equation}
for $t>0$. 

The contribution associated to the positive and negative mode frequencies $u_{\omega}^{(in/out)}$ can be quantized by direct following the standard canonical quantization, where the operators $a_{\omega}^{(in/out)}$ and $a_{\omega}^{(in/out)\dagger}$ act as annihilation and creation operators, respectively. The vacua states $|0_{in}\rangle$ and $|0_{out}\rangle$ are defined as the state with no particles, satisfying
\begin{equation}
    a_{\omega}^{(in/out)}|0_{in/out}\rangle=0,\quad \forall \omega.
\end{equation}
The corresponding excited states (free particles) are then constructed by successive applications of the creation operator on the vacuum, leading to the $n$-particle state
\begin{equation}
|n_{\omega}^{in/out}\rangle=\frac{a_{\omega}^{\dagger}\dots a_{\omega}^{\dagger}}{\sqrt{n!}}|0_{in/out}\rangle.
\end{equation}
These states (including all $n$-particle excitations) span the Fock spaces $\mathscr{F}_{in/out}$, which are the spaces of all possible {\it free}-particles.

However, for the unstable contribution, the decomposition into positive and negative frequency components becomes ill-defined, and the standard quantization procedure is no longer applicable. In other words, there are no natural states in Hilbert space that can span a Fock space, $\mathscr{F}_{\gamma}$, for the {\it unstable}-particles. To address this issue, we first avoid explicitly solving the time component of the unstable mode by expressing $u_{\gamma}(t,x) = \sqrt{2\gamma}e^{-\gamma x}\chi(t)$, and subsequently substituting this expression back into the action~(\ref{action}). This yields
\begin{equation}
       S[u_{\gamma}]=\int_{t_1}^{t_2} dt\left(\frac{\dot{\chi}^2}{2}+\frac{\gamma^2\chi^2}{2}\right),
\end{equation}
which shows that the time evolution of the unstable mode is governed by the equation of an Inverted Harmonic Oscillator (IHO)—a system where the harmonic oscillator's frequency is replaced by $\omega \to \pm i \gamma$. In this analogy, $\chi \in \mathbb{R}$ represents the displacement variable of the oscillator (analogous to the usual $x$-coordinate), and $\dot{\chi}$ corresponds to its conjugate momentum. 

Based on this observation, the framework introduced in Ref.~\cite{felipe-pitelli} provides a method for quantizing the unstable mode by establishing an analogy with the IHO. In the following section, we demonstrate how this approach can be applied to the present case of a single unstable mode.

\section{Unstable mode quantization}\label{sec:quantization}

In quantum mechanics, systems with continuous spectra and non-normalizable eigenstates, such as the inverted harmonic oscillator, present challenges that necessitate an extension of the standard Hilbert space formalism. The IHO is characterized by a parabolic potential, $V(\chi) = -\gamma^2\chi^2 / 2$, which gives rise to scattering dynamics rather than trapped states. To rigorously describe such states, the rigged Hilbert space formalism of quantum mechanics is employed, providing a mathematical framework that incorporates generalized eigenvectors to represent resonances and scattering phenomena. In the following, we review the essential features of this formalism in the quantization of IHO and the unstable mode.

\subsection{Rigged Hilbert space}

The rigged Hilbert space~\cite{bohm, bollini, delamadrid, antoine} is defined by a triplet of spaces (also known as Gelfand triplet):
\begin{equation}\label{RHS}
    \boldsymbol{\Phi} \subset \mathscr{H} \subset \boldsymbol{\Phi}^{\times},
\end{equation}
where $\mathscr{H}$ is the standard infinite dimensional Hilbert space, $\boldsymbol{\Phi}$ is a dense subspace of $\mathscr{H}$ equipped with a nuclear topology, and $\boldsymbol{\Phi}^{\times}$ is the dual space of $\boldsymbol{\Phi}$, consisting of continuous linear functionals. Precisely, a state $\psi$ is a functional belonging to $\boldsymbol{\Phi}^{\times}$ acting as a map from $\boldsymbol{\Phi}$ into the set of complex numbers $\mathbb{C}$, i.e., $\psi:\boldsymbol{\Phi}\to\mathbb{C}$. The triplet (\ref{RHS}) arises from equipping a Hilbert space $\mathscr{H}$ with a topology $\tau_{\mathscr{H}}$, then identifying a subspace $\boldsymbol{\Phi}$ with a stronger topology $\tau_{\boldsymbol{\Phi}}$ than $\tau_{\mathscr{H}}$, and defining the dual space $\boldsymbol{\Phi}^{\times}$ with a topology $\tau_{\boldsymbol{\Phi}^{\times}}$ weaker than $\tau_{\mathscr{H}}$. 

In this framework, one can extend the action of any operator $A$ initially defined on $\mathscr{H}$ to the dual space through the duality formula:
\begin{equation}
    \langle A \varphi | \psi \rangle = \langle \varphi | A^{\times} \psi \rangle, \quad \forall \varphi \in \boldsymbol{\Phi},\, \forall \psi \in \boldsymbol{\Phi}^{\times},
\end{equation}
where $A^{\times}$ denotes the extension of $A$ to the larger space $\boldsymbol{\Phi}^{\times}$. The RHS accommodates the Dirac bracket formalism (see Refs.~\cite{delamadrid, bohm-gadella-mainland}) by interpreting the {\it kets} as elements of the largest space and the {\it bras} as elements of the smallest space, ensuring that the bracket operation converges\footnote{In the standard Hilbert space formulation of quantum mechanics, due to the Riesz–Fischer theorem~\cite{dunford-schwartz}, one has the isometry $\mathscr{H}\simeq\mathscr{H}^{\times}$, and therefore it is not necessary to distinguish between the spaces of bras and kets.}. We will adopt these definitions for bras and kets in the rest of this work.

A complex number $\lambda$ is a {\it generalized eigenvalue} of $A$ if there exists a non-zero $\psi \in \boldsymbol{\Phi}^{\times}$ such that for all $\varphi \in \boldsymbol{\Phi}$,
\begin{equation}\label{duality-relation}
    \langle A\varphi|\psi\rangle=\langle\varphi|A^{\times}\psi\rangle=\lambda\langle\varphi|\psi\rangle.
\end{equation}
If we omit the arbitrary $\varphi$, we have the generalized eigenvalue equation
\begin{equation}
    A^{\times}|\psi\rangle=\lambda|\psi\rangle,
\end{equation}
with $\lambda$ being a generalized eigenvalue of $A$, which can be either continuous or complex, as the generalized eigenvector $\psi$ belongs to the larger space $\boldsymbol{\Phi}^{\times}$. In other words, the RHS provides a natural extension of quantum theory to handle continuous spectra, with its mathematical foundation established by the Nuclear Spectral Theorem~\cite{maurin}.

Now we are ready to quantize the unstable mode as an inverted harmonic oscillator within the RHS formalism. To this end, we must analyze two distinct bases of the IHO: the energy eigenstates and the decay/growth states.

\subsection{Energy eigenstate}

The unstable mode is governed by the Hamiltonian
\begin{equation}
    H=\frac{\dot{\chi}^2}{2}-\frac{\gamma^2\chi^2}{2},
\end{equation}
with the quantization condition $[\chi,\dot{\chi}]=i(\hslash=1)$. This Hamiltonian exhibits a twofold degeneracy in its generalized eigenvalues $E \in \mathbb{R}$, given by
\begin{equation}
H^{\times}|\psi^{E}_{\pm}\rangle=E|\psi^{E}_{\pm}\rangle.
\end{equation}

In the ``position'' $\chi$ representation, $\psi^{E}_{\pm}(\chi)=\langle \chi|\psi^{E}_{\pm}\rangle$, the general solution is expressed as a linear combination of parabolic cylinder functions that satisfy some appropriate boundary condition. For our purpose, we choose the solution (see Ref.~\cite{chruscinski} for details)
\begin{equation}\label{energy1}
    \psi^{E}_{\pm}(\chi) = \frac{C_0}{\sqrt{2\pi \gamma}} i^{\frac{\nu+1}{2}} \Gamma(\nu+1) D_{-\nu-1}(\mp \sqrt{-2i\gamma}\chi),
\end{equation}
where 
\begin{equation*}
    \nu = -\left( i \frac{E}{\gamma} + \frac{1}{2} \right)\quad \text{and}\quad C_0 = \left( \frac{\gamma}{2\pi^2} \right)^{1/4}.
\end{equation*}
These eigenfunctions are incomplete without the conjugate solutions $\overline{\psi^{E}_{\pm}}$, which satisfy the eigenvalue equation $H^{\times}\overline{\psi^{E}_{\pm}}=-E\overline{\psi^{E}_{\pm}}$. Explicitly, we have
\begin{equation}\label{energy2}
    \overline{\psi^{E}_{\pm}(\chi)} = \frac{C_0}{\sqrt{2\pi \gamma}} i^{\frac{\nu+1}{2}} \Gamma(-\nu) D_{\nu}(\mp \sqrt{2i\gamma}\chi).
\end{equation}

The parabolic cylinder function, $D_{\nu}$, exhibits exponential divergence as $|\chi| \to \infty$, and therefore, (\ref{energy1}) and (\ref{energy2}) are not square-integrable, meaning they do not belong to the Hilbert space $\mathscr{H}=\mathcal{L}^2(\mathbb{R}_{\chi})$. However, they satisfy orthogonality relations in the sense of distributions:
\begin{equation}
    \begin{aligned}
        &\int_{\mathbb{R}} \overline{\psi^{E}_{\pm}(\chi)} \psi^{E^{\prime}}_{\pm}(\chi) \, d\chi = \delta(E-E^{\prime})\\
        &\int_{\mathbb{R}} \overline{\psi^{E}_{\pm}(\chi)} \psi^{E}_{\pm}(\chi^{\prime}) \, dE = \delta(\chi-\chi^{\prime}).
    \end{aligned}
\end{equation}
In this way, the appropriate choice for defining the RHS of the inverted harmonic oscillator is to set $\boldsymbol{\Phi}$ as the Schwartz space $\mathcal{S}(\mathbb{R}_{\chi})$ — the space of rapidly decreasing functions that vanish faster than any inverse polynomial as $|\chi|\to\infty$. Consequently, its dual, $\mathcal{S}(\mathbb{R}_{\chi})^{\times}$, is the space of distributions, in which the eigenvectors of the IHO are well-defined. Thus, the RHS of the IHO is characterized by the triplet
\begin{equation}
    \mathcal{S}(\mathbb{R}_{\chi})\subset\mathcal{L}^2(\mathbb{R}_{\chi})\subset\mathcal{S}^{\times}(\mathbb{R}_{\chi}).
\end{equation}

\subsection{Decay and growth states}

Similar to the standard harmonic oscillator, ladder operators can be introduced as
\begin{equation}\label{b+- definition}
    b^{\pm} := \sqrt{\frac{\gamma}{2}} \left( \chi \mp \frac{\dot{\chi}}{\gamma} \right).
\end{equation}
In this basis, the commutation relations writes
\begin{equation}
    [b^{+}, b^{-}] = i, \quad [b^{\pm}, b^{\pm}] = 0.
\end{equation}

The (unstable mode) Hamiltonian can be expressed as
\begin{equation}
    H = -\frac{\gamma}{2}(b^{+}b^{-} + b^{-}b^{+})
\end{equation}
and the operators $b^{\pm}$ are extended to a larger space by the duality formula (\ref{duality-relation}), satisfying the condition $(b^{\pm})^{\times}f=b^{\pm}f$ for any $f\in\boldsymbol{\Phi}^{\times}$, where $\boldsymbol{\Phi}=\mathcal{S}(\mathbb{R}_{\chi})$. The ground states $|f_0^{\pm}\rangle$ are then defined by
\begin{equation}
    b^{\pm}|f_0^{\mp}\rangle= 0,
\end{equation}
indicating that $b^+$ annihilates $f_0^-$, while $b^-$ annihilates $f_0^+$. The excited states are constructed as
\begin{equation}
    |f_n^{\pm}\rangle = (b^{\pm})^n |f_0^{\pm}\rangle, \quad n \in \mathbb{Z}^+_0,
\end{equation}
and correspond to eigenvectors of the Hamiltonian with purely imaginary eigenvalues:
\begin{equation}\label{hamiltonian2}
    H^{\times}|f_n^{\pm}\rangle = \pm E_n |f_n^{\pm}\rangle, \quad E_n = i\gamma \left( n + \frac{1}{2} \right).
\end{equation}

In the $\chi$ representation, the solution to Eq.~(\ref{hamiltonian2}) is analogous to the solution of the standard harmonic oscillator setting $\omega\to\pm i\gamma$:
\begin{equation}\label{fn-position}
    f_n^{\pm}(\chi) = N_n^{\pm} e^{\mp \frac{i\gamma \chi^2}{2}} H_n(\sqrt{\pm i\gamma} \chi),
\end{equation}
where $H_n$ are Hermite polynomials and $N_n^{\pm}$ are normalization constants. The states described by (\ref{fn-position}) are generalized states that characterize the system, with the following properties:
\begin{enumerate}
    \item {\it Conjugacy}\label{property1} - These states are conjugate to each other:
        \begin{equation}
            \overline{f_n^+(\chi)}=f_n^-(\chi).
        \end{equation}
    \item {\it Orthonormality}\label{property2} - They are orthonormal:
        \begin{equation}
            \langle f_{n}^{\pm}|f_{n}^{\mp}\rangle=\delta_{nm}.
        \end{equation}
    \item {\it Completeness}\label{property3} - They form a complete basis:
        \begin{equation}
            \sum_{n=0}^{\infty}\overline{f_n^{\pm}(\chi)}f_n^{\mp}(\chi^{\prime})=\delta(\chi-\chi^{\prime}).
        \end{equation}
\end{enumerate}

Notably, property \ref{property2} implies the {\it bra} element corresponding to the {\it ket} $|f_n^{\pm}\rangle$ is $\langle f_n^{\mp}|$, i.e., the sign in the notation switches from $+$ to $-$ (and vice versa). Moreover, property 3 highlights that these states do not belong to the Hilbert space $\mathscr{H} = \mathcal{L}^2(\mathbb{R}_{\chi})$. To distinguish between the spaces of $f_n^+$ and $f_n^-$, we introduce two triplets of spaces:
\begin{equation}\label{triplet22}
    \boldsymbol{\Phi}_{\pm}\subset\mathscr{H}\subset\boldsymbol{\Phi}_{\pm}^{\times},
\end{equation}
where $|f_n^{+}\rangle\in \boldsymbol{\Phi}_{+}^{\times}$ and $|f_n^{-}\rangle\in \boldsymbol{\Phi}_{-}^{\times}$, with $\boldsymbol{\Phi}_{+}\cap\boldsymbol{\Phi}_{-}=\{\emptyset\}$.

The spaces $\boldsymbol{\Phi}_{\pm}$ were rigorously defined by Chruściński~\cite{chruscinski,chruscinski-jmp}. By analyzing the analytic properties of $\psi_{\pm}^{E}$ and $\overline{\psi_{\pm}^{E}}$, it is evident that these functions have simple poles at $E=-E_n$ and $E=E_n$. When analytically continued to the complex plane, these poles correspond to complex energies located on the imaginary axis, with the residues defining the states $f_n^{\pm}$ (see also Ref.~\cite{marcucci-conti}). This provides a systematic definition for the spaces of decay and growth states as
\begin{equation}\label{space-definition}
\begin{aligned}
      \boldsymbol{\Phi}_{-}&=\left\{\varphi \in \mathcal{S}(\mathbb{R}_{\chi}) \,\big| \langle\varphi|\overline{\psi_{\pm}^{E}}\rangle\in \mathcal{S}(\mathbb{R}_{E})\cap\mathscr{H}^{2}_{-}(\mathbb{R}_{E})\right\}\\
       \text{and} &\\
       \boldsymbol{\Phi}_{+}&=\left\{\varphi \in \mathcal{S}(\mathbb{R}_{\chi}) \,\big|\langle\varphi|\psi_{\pm}^{E}\rangle\in \mathcal{S}(\mathbb{R}_{E})\cap\mathscr{H}^{2}_{+}(\mathbb{R}_{E})\right\},
\end{aligned}
\end{equation}
where $\mathscr{H}_{+}^2(\mathscr{H}_{-}^2)$ denotes the Hardy class space~\cite{duren} for the upper (lower) half-plane. Specifically, $\boldsymbol{\Phi}_{+}$ consists of functions that are boundary values of analytic functions in the upper half-plane of the complex $E$-plane and vanish faster than any power of $E$ along the upper semicircle. Similarly, $\boldsymbol{\Phi}_{-}$ corresponds to the analogous space for functions analytic in the lower half-plane.

Having established these functional spaces and their analytic character, we can now examine how they influence the dynamical description of the unstable sector.

\subsection{Time evolution}

As the unstable mode is characterized by an imaginary energy, the decaying and growing states define the natural ket states to describe the corresponding unstable particle. Due to the newly defined basis, the two spaces introduced in (\ref{space-definition}) impose constraints on operators previously defined in the Hilbert space $\mathscr{H}$. One prominent example is the unitary time evolution operator $U(t) = e^{-i H t}$. For $\varphi^{+}\in\boldsymbol{\Phi}_{+}$, the action of $U(t)$ on $\boldsymbol{\Phi}_{+}$ is given by
\begin{equation}\label{exp1}
    \begin{aligned}
        \langle U(t)\varphi^{+}|\psi_{\pm}^{E}\rangle&=\langle\varphi^{+}|U^{\times}(t)\psi_{\pm}^{E}\rangle\\
        &=e^{iEt}\langle\varphi^{+}|\psi_{\pm}^{E}\rangle\,\in\,\mathscr{H}_{+}^{2}\,\Leftrightarrow \,t\geq 0.
    \end{aligned}
\end{equation}
That is, the state $\varphi^{+}$ evolves in time only for $t> 0$ when constrained into the dense subspace $\boldsymbol{\Phi}_{+}$. Similarly, for $\varphi^{-}\in\boldsymbol{\Phi}_{-}$ we find
\begin{equation}\label{exp2}
\begin{aligned}
    \langle U(t)\varphi^{-}|\overline{\psi_{\pm}^{E}}\rangle&=\langle\varphi^{-}|U^{\times}(t)\overline{\psi_{\pm}^{E}}\rangle\\
    &=e^{-iEt}\langle\varphi^{-}|\overline{\psi_{\pm}^{E}}\rangle\,\in\,\mathscr{H}_{-}^{2}\,\Leftrightarrow \,t\leq 0.
\end{aligned}
\end{equation}

Thus, the operator $U(t)$ naturally splits into two semigroups:
\begin{equation} \begin{aligned} 
&U_{+}(t) = U(t)\big|_{\boldsymbol{\Phi}_{+}}: \boldsymbol{\Phi}_{+} \to \boldsymbol{\Phi}_{+} \quad \textrm{for}\quad t \geq 0,\\ 
&U_{-}(t) = U(t)\big|_{\boldsymbol{\Phi}_{-}}: \boldsymbol{\Phi}_{-} \to \boldsymbol{\Phi}_{-} \quad \textrm{for}\quad t \leq 0. \end{aligned} \end{equation} 
This splitting reflects the intrinsic time asymmetry of the system---namely, the unstable mode is not invariant under time reversal ($t\to-t$). In particular, the vectors $f_n^{+} \in \boldsymbol{\Phi}_{+}^{\times}$ are restricted to exist for $t \geq 0$, while $f_n^{-} \in \boldsymbol{\Phi}_{-}^{\times}$ exist only for $t \leq 0$. This temporal distinction naturally assigns the labels ``decay states'' to $f_n^{+}(t) = U^{\times}_{+}(t) f_n^+$ and ``growth states'' to $f_n^{-}(t) = U^{\times}_{-}(t) f_n^{-}$, reflecting their dynamical behavior over time.

In field theory, we typically work in the Heisenberg picture, where field operators evolve in time while the states remain fixed. For the operators $b^{\pm}$, their time evolution are given by
\begin{equation}
    \begin{aligned}
        b^{\pm}(t)&=U_{\pm}^{\times}(t)b^{\pm}U_{\pm}(t)\\ &\Rightarrow b^{\pm}(t)=b^{\pm}e^{\mp \gamma t},\quad \pm t\geq 0,
    \end{aligned}
\end{equation}
with initial condition $b^{\pm}(0)=b^{\pm}$. Substituting this result into Eq.~(\ref{b+- definition}), we can express the unstable mode as
\begin{equation}\label{u-gamma}
    u_{\gamma}=b^+u^{(+)}+b^-u^{(-)},
\end{equation}
where
\begin{equation}
    u^{(\pm)}(t,x)=\exp(-\gamma x\mp \gamma t),\quad \pm t\geq 0,
\end{equation}
are the natural ``modes'' describing the decaying and growing components of the unstable particle.

At this point, we have both a physical and a mathematical justification for selecting either the exponentially decaying or growing solution. In our case, where the Robin boundary condition is imposed only for $t > 0$, the appropriate choice is $u_{\gamma} = b^{+} u^{(+)}$, which corresponds to the surviving contribution of the unstable mode in the domain $t \geq 0$. Notably, this solution no longer diverges for large values of $t$; that is, at the quantum level, the RHS formalism absorbs the classical divergence into a consistent decaying quantum state. Furthermore, under time reversal $t \to -t$, the mode transforms as  $u^{(+)} \to u^{(-)}$, further illustrating the time asymmetry of the unstable mode—an effect naturally captured within the rigged Hilbert space framework for time-asymmetric systems~\cite{bohm-time-asymmetric}.

\section{Unruh–DeWitt detector coupled to an unstable mode}\label{sec:detector}

To describe the interaction between an observer and the unstable mode, we consider the interaction Hamiltonian of an Unruh–DeWitt detector coupled to the total field. This Hamiltonian splits into two contributions: one from the free field, $H_{\phi_f}$, and another from the unstable mode, $H_{u_{\gamma}}$, as follows:
\begin{equation}\label{interaction-hamiltonian}
    H_{\text{int}}=H_{\phi_f}+H_{u_{\gamma}},
\end{equation}
where
\begin{equation}
   H_{\phi_f}=\lambda\,\eta(\tau)\mu(\tau)\phi_f[{\bf x}(\tau)],
\end{equation}
and
\begin{equation}
    H_{u_{\gamma}}=\lambda\,\eta(\tau)\mu(\tau)u_{\gamma}[{\bf x}(\tau)].
\end{equation}
Here, $\phi_f$ denotes the free-field component, defined by Eq.~\eqref{before} for $t < 0$ and by the integral contribution in Eq.~\eqref{after} for $t>0$. For instance, in order to develop a general framework, let's consider $u_{\gamma}$ as given by Eq.~\eqref{u-gamma}.

In the equations above, $\lambda$ is a small coupling constant, $\eta(\tau)$ is the switching function that controls the interaction time duration $T$, ${\bf x}(\tau)=(t(\tau),x(\tau))$ is the detector trajectory, $\tau$ is the detector's proper time, and $\mu(\tau)$ is the monopole momentum operator, which mediates transitions between the detector's two-level states as
\begin{equation}
    \mu(\tau)=|e\rangle\langle g|e^{i\Omega \tau}+|g\rangle\langle e|e^{-i\Omega \tau},
\end{equation}
with $|g\rangle$ and $|e\rangle$ denoting the detector's ground and excited states, respectively. $\Omega$ is the energy gap between these states.

The detector's Hilbert space is $\mathcal{H}_D \cong \mathbb{C}^2$, spanned by the states $\{|g\rangle,|e\rangle\}$. The Hilbert space of the stable sector is the Fock space $\mathscr{F}_{in/out}$, generated by the vacua $|0_{in/out}\rangle$. The unstable sector is characterized by the RHS \eqref{triplet22}, so that the states $\{|f_n^{\pm}\rangle\}$ span $\boldsymbol{\Phi}_{\pm}^{\times}$. Accordingly, the total Hilbert space of the system is given by $\mathcal{H}_D\otimes\mathscr{F}_{in}$ for $t<0$, and by $\mathcal{H}_D\otimes\mathscr{F}_{out}\otimes\boldsymbol{\Phi}^{\times}_{+}$ for $t\geq0$.

For simplicity, we assume that the system is initially in its respective ground states: the detector begins in $|g\rangle$, while the free field and the unstable mode start in their respective vacuum states, $|0_{in/out}\rangle$ and $|f_0^{+}\rangle$. The unstable mode depends on the proper time $\tau$, which may select a positive or negative time coordinate. Specifically, for $t(\tau)< 0$, there is no unstable mode, whereas for $t(\tau)> 0$ it is $|f_0^{+}\rangle$. After the interaction, the detector exchanges energy $\Omega$ with the field, resulting in a transition of the detector to $|e\rangle$. For the free field, this corresponds to the state $|1_{\omega}^{in/out}\rangle$, and for the unstable mode, to $|f_n^{+}\rangle$.

By applying perturbation theory, the transition amplitude for  $|g\rangle\otimes|0_{in/out}\rangle\otimes|f_0^{+}\rangle\to|e\rangle\otimes|1_{\omega}^{in/out}\rangle\otimes |f_n^{+}\rangle$ is given by (at first order in $\lambda$)
\begin{equation}\label{amplitude}
    \mathcal{A}=\mathcal{A}_{\phi_f}+\mathcal{A}^{+},
\end{equation} 
where
\begin{equation}\label{amplitude-f}
    \mathcal{A}_{\phi_f}=-i\lambda\int d\tau e^{i\Omega \tau}\eta(\tau)\langle 1_{\omega}^{in/out}|\phi_f[{\bf x}(\tau)]|0_{in/out}\rangle\langle f_n^{-}|f_0^{+}\rangle,
\end{equation}
and 
\begin{equation}\label{amplitude+-}
    \mathcal{A}^{+}=-i\lambda \int d\tau e^{i\Omega \tau}\eta(\tau)\langle 1_{\omega}^{in/out}|0_{in/out}\rangle\langle f_n^{-}|u_{\gamma}[{\bf x}(\tau)]|f_0^{+}\rangle.
\end{equation}
Note that the {\it bra} element is written with the sign exchange $+ \to -$, as required by property \ref{property2}. Analyzing these expressions, we observe that the matrix element $\langle f_n^{-}|u_{\gamma}|f_0^{+}\rangle$ imposes a constraint on the time domain of $u_{\gamma}$, i.e., even if we consider $u_{\gamma}$ as given by Eq.~\eqref{u-gamma}, the {\it bracket} operation constrain the time domain. Explicitly, this term gives us
\begin{equation}\label{expected-value}
    \langle f_n^{-}|u_{\gamma}({\bf x})|f_0^{+}\rangle=u^{(+)}({\bf x})\delta_{n1},\quad t(\tau)\geq 0,
\end{equation}
which is identically zero for $ t(\tau)<0$.

Interestingly, the results (\ref{amplitude})--(\ref{amplitude+-}) show that, at first order in $\lambda$, if we consider transitions only for $t\geq 0$, there is no \textit{backreaction} between the contributions of the free field and the unstable mode, i.e., the free field depositing energy in the detector and the detector transferring that energy into the unstable field and vice versa. In other words, if the detector is initialized in a pure state, it will exchange a real energy $\Omega$ with either the free field or the unstable mode, but not both ``simultaneously'', as the amplitude elements are off-diagonal.

Finally, taking the modulus square of Eq.~(\ref{amplitude}) and summing over all possible final states $|1_{\omega}^{in/out}\rangle$, we obtain the transition probability:
\begin{widetext}
\begin{equation}\label{probability}
         \mathcal{P}^{+}(\Omega)=\lambda^2\int_{I_{+}} d\tau \int_{I_{+}} d\tau^{\prime}\eta(\tau)\eta(\tau^{\prime})e^{-i\Omega(\tau-\tau^{\prime})}e^{-\gamma\left[x(\tau)+x(\tau^{\prime})\right]}e^{- \gamma \left[t(\tau)+t(\tau^{\prime})\right]},
\end{equation}
\end{widetext}
where $I_{+} := \{\, \tau \in \mathbb{R} \mid t(\tau) \geq 0 \,\}$.

This expression encapsulates the interaction between the detector and the unstable mode. Since our primary objective is to isolate and examine the effects associated with the unstable dynamics, we shall henceforth focus exclusively on the probability given in Eq.~\eqref{probability}, disregarding any contributions from the free field---for a detailed discussion of the free-field contribution in a conformally related setting, see Ref.~\cite{pitelli-barroso}. We now proceed to evaluate the response of the detector by specifying its worldline and analyzing how the interaction with the unstable mode unfolds along different types of trajectories.
\begin{figure}[!htb]
    \centering
    \includegraphics[width=0.35\linewidth]{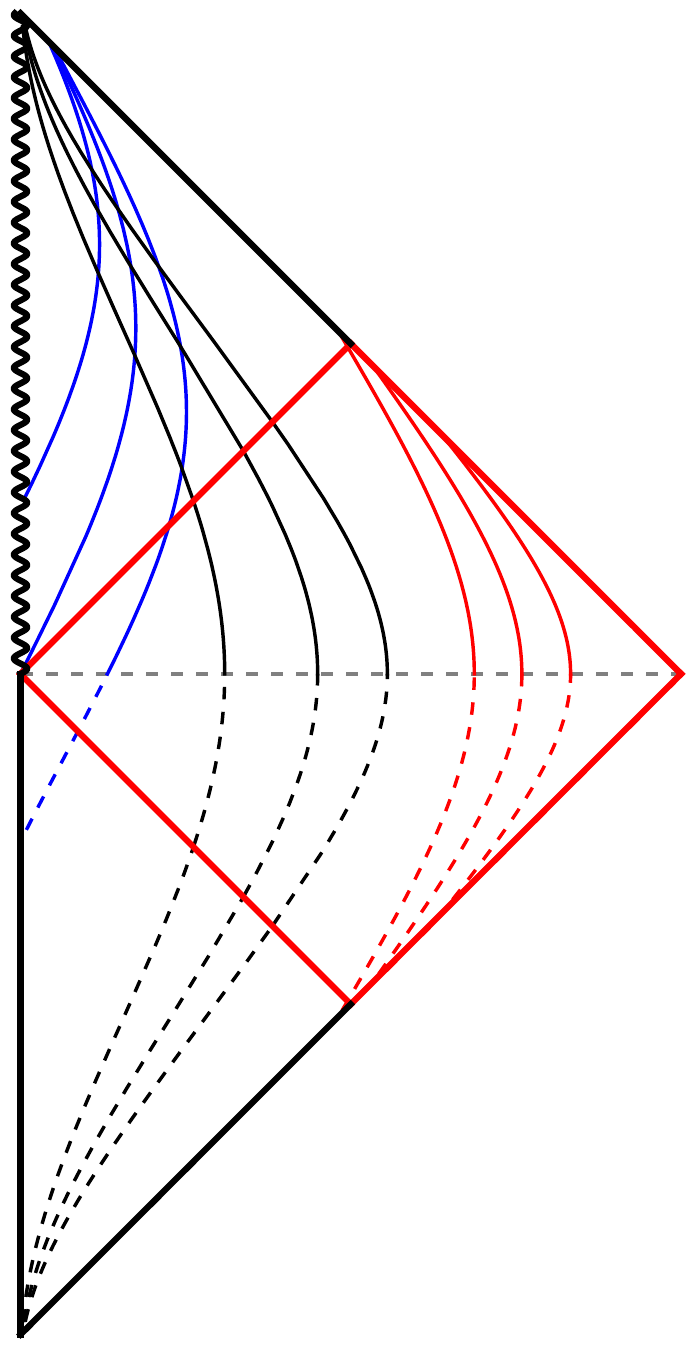}
    \caption{Conformal diagram illustrating static (black), inertial (blue), and uniformly accelerated (red) detector trajectories in (1+1)-dimensional half-Minkowski spacetime. Solid lines indicate regions where the detector interacts with the unstable mode ($t>0$).}
    \label{fig:trajectories}
\end{figure}

In Fig.~\ref{fig:trajectories}, black curves depict static observers at fixed spatial positions $x_0$. Blue curves represent inertial observers with constant velocity $v$, starting from position $x_0$. Red curves, confined to the diamond-shaped Rindler wedge, correspond to the path of uniformly accelerated observers parametrized by a proper acceleration $a$. After $t = 0$, all trajectories are shown as solid lines, indicating the region of the trajectory where the detector interacts with an unstable mode.

\subsection{Static trajectory}

The simplest detector trajectory to consider is one in which it remains at rest at a fixed position $x_0$, evolving solely in time (see Fig.~\ref{fig:trajectories}). This case is particularly relevant because the trajectory is parametrized by the coordinate time $t$, following the integral curves of the timelike Killing vector field $\partial_t$. As such, it represents the ``natural'' path for laboratory observers.

In terms of the detector’s proper time, this trajectory is given by
\begin{equation}\label{trajectory}
   t(\tau)=\tau, \quad x(\tau)=x_0.
\end{equation}

Assuming a sharp switching function of the form $\eta(\tau)=\Theta(T-\tau)$, and taking the limit of an eternally switched-on detector $T\to \infty$, the transition probability \eqref{probability} for a static observer becomes
\begin{equation}
    \mathcal{P}^{+}=\lambda^2e^{-2\gamma x_0}\int_{0}^{\infty}d\tau\int_{0}^{\infty}d\tau^{\prime}e^{-i\Omega(\tau-\tau^{\prime})}e^{-\gamma (\tau+\tau^{\prime})},
\end{equation}
which can be directly evaluated to yield
\begin{equation}\label{probability2}
    \mathcal{P}^{+}(\Omega)=\lambda^2e^{-2\gamma x_0}\frac{1}{\Omega^2+\gamma^2}.
\end{equation}

At $x_0=0$, the result in Eq.~\eqref{probability2} takes the form of an unnormalized Breit-Wifner distribution~\cite{breit-wigner}
\begin{equation}\label{BW}
    f(E)\sim \frac{1}{(E-E_0)^2+\Gamma_0^{2}/4},
\end{equation}
which characterizes the energy spectrum $E$ of an unstable state. This profile encapsulates the main features of a resonance: the peak at $E_0$ corresponds to the most probable energy, while the width $\Gamma_0$ determines the finite mean lifetime $\tau \sim 1/\Gamma_0$. The Breit–Wigner distribution thus establishes the natural connection between exponential decay in time and the spectral representation of unstable states. In the present case, comparing Eqs.~\eqref{probability2} and \eqref{BW} identifies $\Omega: = E - E_0$ and $\Gamma _0:= 2\gamma$. Consequently, a detector at rest registers a resonant transition probability peaked at $\Omega=0$, with a decay width proportional to the boundary parameter $\gamma$. As the detector is placed farther from the singularity, the amplitude of this resonance is exponentially suppressed by the prefactor $e^{-2\gamma x_0}$, reflecting the localized nature of the unstable mode near the boundary.

The combined system of the detector and the unstable mode can then be regarded as a resonant system, where the imaginary energy of the unstable mode determines the mean lifetime $\tau_{\text{mean}}\sim1/2\gamma$. As  $\gamma \to \infty$  (corresponding to the Dirichlet boundary condition), the decay width diverges, leading to $\mathcal{P}^{+}(\Omega) \to 0$. This aligns with the fact that Dirichlet conditions suppress the existence of unstable modes altogether. Conversely, in the Neumann limit $\gamma \to 0$, the decay width vanishes, and the mean lifetime becomes infinite, i.e., the detector interact with the non-oscillatory (zero mode) particle allowed by the Neumann boundary condition \footnote{For a complete analysis of the non-oscillatory zero mode interacting with an Unruh-DeWitt detector, we refer to \cite{loko}}. In this case, the transition probability becomes $\mathcal{P}^{+}(\Omega) =\lambda^2 / \Omega^2$, exhibiting an energy divergence in the infrared limit. 

This result shows that the quantization of the unstable state within the RHS formalism yields a well-defined decaying state, characterized by the Breit–Wigner spectral profile---the standard probability distribution for decay and resonant systems~\cite{weinberg,cohen}. Having understood this connection, we now consider inertial motion to examine how constant velocity influences the decay structure.

\subsection{Inertial trajectory}

To investigate how the detector’s kinematics influences its response to the unstable mode, we now consider inertial trajectories with constant velocity $v<1$. In this case, the detector moves along the worldline:
\begin{equation}\label{supercritical}
t(\tau) = \frac{1}{\sqrt{1-v^2}}\tau, \quad x(\tau) = x_0+\frac{v}{\sqrt{1-v^2}}\tau,
\end{equation}
where $\tau$ is the proper time along the detector’s path and $x_0$ denotes the initial position at $\tau = 0$.

Substituting this trajectory into Eq.~\eqref{probability}, and again assuming a sharp switching function, the transition probability becomes
\begin{equation}
    \mathcal{P}^{+}(\Omega)=\lambda^2 e^{-2\gamma x_0}\int_{0}^{\infty}d\tau \int_{0}^{\infty} d\tau^{\prime}e^{-i\Omega(\tau-\tau^{\prime})}e^{-\gamma \frac{(v+1)(\tau+\tau^{\prime})}{\sqrt{1-v^2}}},
\end{equation}
yielding after integration:
\begin{equation}\label{probability3}
    \mathcal{P}^{+}(\Omega)=\lambda^2 e^{-2\gamma x_0}\frac{1}{\Omega^2+\left(\frac{1+v}{1-v}\right)\gamma^2}.
\end{equation}

In the result above, we observe that the detector’s velocity $v$ modifies the decay width $\Gamma_0$, while the asymptotic behavior of the boundary parameter $\gamma$---as it approaches the Dirichlet or Neumann boundary conditions---remains qualitatively the same as in the static case. As a consistent check, note that in the limit $v\to 0$, Eq.~\eqref{probability3} reduces to Eq.~\eqref{probability2}, thus recovering the static scenario as a particular case. 

By comparing Eq.~\eqref{probability3} to the Breit–Wigner forms, we identify the decay width as
\begin{equation}
    \Gamma_v:=2\gamma \sqrt{\frac{1+v}{1-v}}.
\end{equation}
This expression shows that $\Gamma_v$ increases monotonically with $v$ and diverges as $v \to 1$ by a {\it relativistic Doppler factor}, where $\Gamma_0$ is the ``emitted frequency.'' Physically, this implies that the mean lifetime $\tau_{\text{mean}}\sim 1/\Gamma_v$ decreases as the detector’s velocity increases. In the ultrarelativistic limit $v \to 1$, the decay becomes instantaneous and the detector registers no transition---indicating that the interaction with the unstable mode is suppressed in this regime.

\begin{figure}[!htb]
    \centering
    \includegraphics[width=.9\linewidth]{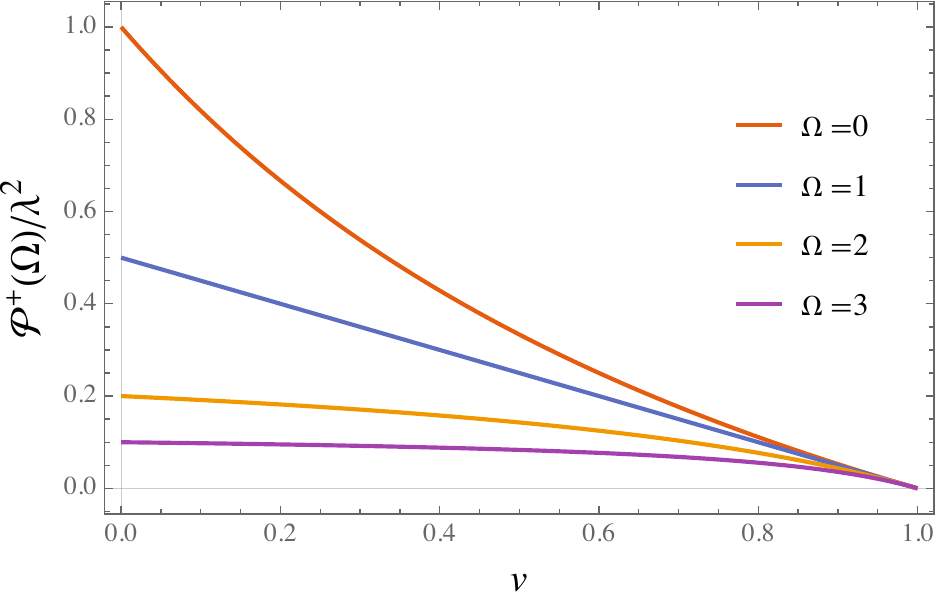}
    \caption{Transition probability $\mathcal{P}^{+}(\Omega)$ over $\lambda^2$, given by Eq.~\eqref{probability3}, as a function of detector velocity $v$. We consider $x_0 = 0$ and $\gamma = 1$. As $v \to 1$, the probability sharply decreases, indicating the suppression of detection in the ultrarelativistic regime.}
    \label{fig:prob1}
\end{figure}

Figure~\ref{fig:prob1} illustrates this behavior by showing the probability $\mathcal{P}^{+}(\Omega)$ as a function of $v$, for fixed parameters.

\subsection{Accelerated trajectory}

We now consider a uniformly accelerated detector, characterized by a constant proper acceleration $a$. Such an observer follows a hyperbolic trajectory confined to the Rindler wedge of Minkowski spacetime, as depicted in Fig.~\ref{fig:trajectories}. The worldline is parametrized by the proper time $\tau$ as
\begin{equation}\label{aceleration}
    t(\tau)=\frac{1}{a}\sinh(a\tau),\quad x(\tau)=\frac{1}{a}\cosh(a\tau).
\end{equation}

In this scenario, the transition probability changes substantially. Substituting the trajectory into Eq.~\eqref{probability} and considering an eternally switched-on detector, we obtain
\begin{equation}
    \mathcal{P}^{+}(\Omega)=\lambda^2\int_{0}^{\infty}d\tau \int_{0}^{\infty}d\tau^{\prime} e^{-i\Omega(\tau-\tau^{\prime})}\exp\left[-\frac{\gamma}{a}(e^{a\tau}+e^{a\tau^{\prime}})\right].
\end{equation}
Using formula 3.331–2 from Ref.~\cite{gradshteyn-ryzhik}, this integral can be evaluated exactly, yielding
\begin{equation}\label{probability4}
    \mathcal{P}^{+}(\Omega)=\lambda^2 \frac{\Gamma\left(-\frac{i\Omega}{a},\frac{\gamma}{a}\right)\Gamma\left(\frac{i\Omega}{a},\frac{\gamma}{a}\right)}{a^2},
\end{equation}
where $\Gamma(\alpha,y)$ denotes the incomplete gamma function.

\begin{figure}[!htb]
    \centering
    \includegraphics[width=0.9\linewidth]{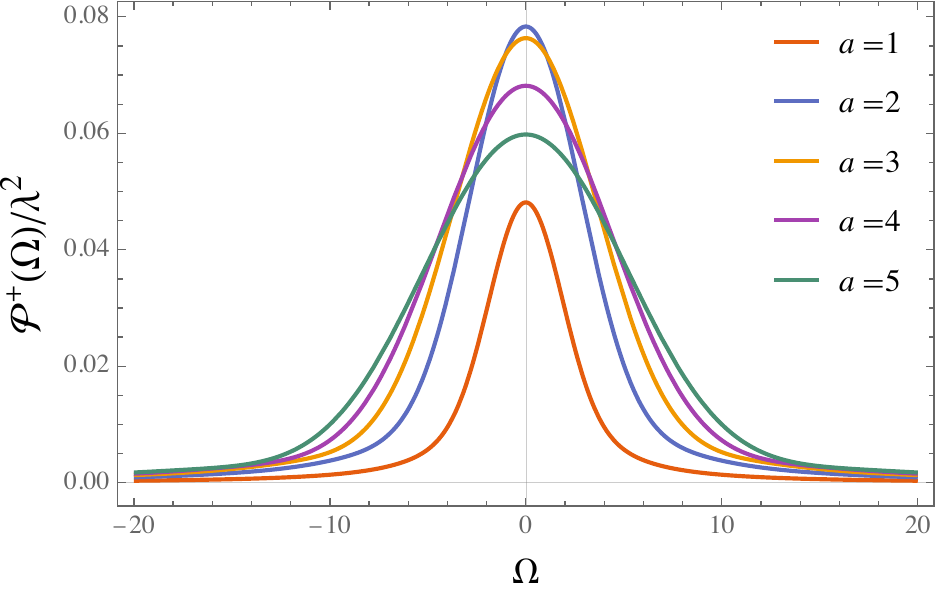}
    \caption{Transition probability $\mathcal{P}^{+}(\Omega)/\lambda^2$ for a uniformly accelerated detector, as given by Eq.~\eqref{probability4}. Here we consider different accelerations for a fixed boundary parameter $\gamma=1$.}
    \label{fig:prob2}
\end{figure}

Unlike the previous cases, the dependence of the transition probability on the energy gap $\Omega$ and acceleration $a$ is no longer of Breit–Wigner form, and its analytical structure becomes considerably more intricate. To explore its behavior, we examine Eq.~\eqref{probability4} numerically. In Fig.~\ref{fig:prob2} we can see $\mathcal{P}^{+}(\Omega)$ as a function of the energy gap for some values of the acceleration. Despite the complexity of the analytical form, the probability remains peaked around $\Omega = 0$, regardless of the specific values of $a$ and $\gamma$, preserving the resonance-like profile observed in inertial cases.

\begin{figure}[!htb]
    \centering
    \includegraphics[width=0.9\linewidth]{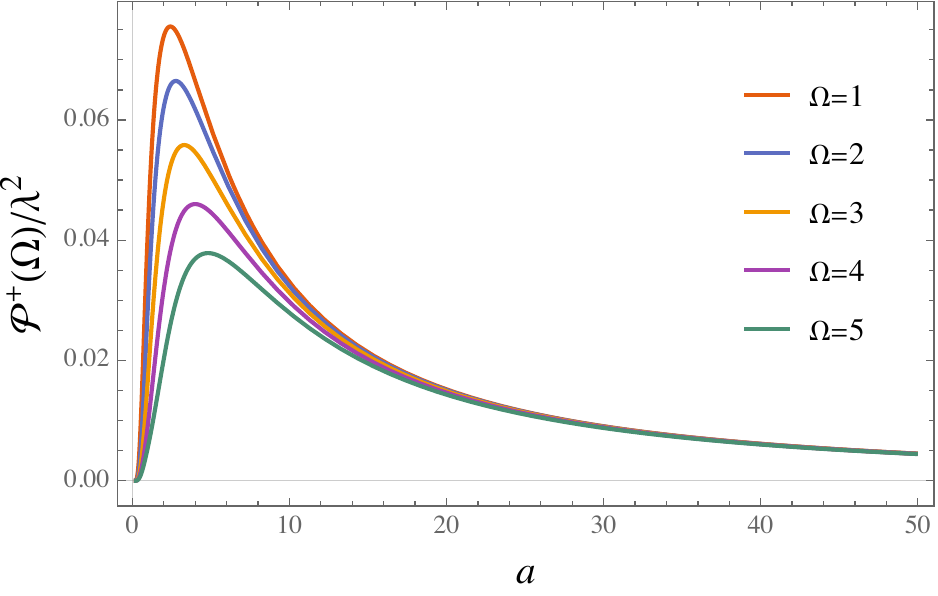}
    \caption{Transition probability $\mathcal{P}^{+}(\Omega)/\lambda^2$ given by Eq.~\eqref{probability4} as a function of the proper acceleration. We are setting $\gamma=1$.}
    \label{fig:prob3}
\end{figure}

We next examine the detector’s response as a function of the proper acceleration $a$. Figure~\ref{fig:prob3} displays the behavior of  $\mathcal{P}^{+}(\Omega)$ for various energy gaps. As observed, the transition probability vanishes at $a = 0$, increases with acceleration, reaches a maximum at a critical value $a_{\max}$, and then decreases toward zero in the limit $a \to \infty$. This is consistent with the idea that ultrarelativistic motion suppresses the detection of unstable particles, while also reflecting the suppression in probability due to a large value of $x$ [note that $x(\tau)\to\infty$ as $a\to0$ in Eq.~\eqref{aceleration}]. The peak positions $a_{\max}$, corresponding to the maximum detector response, can be numerically extracted from the condition $d\mathcal{P}^{+}(\Omega)/da = 0$. For the cases shown in Fig.~\ref{fig:prob3}, we find:
\begin{equation}
    a_{\max}=2.42,\, 2.76,\, 3.30,\, 4.00,\, 4.82,
\end{equation}
for $\Omega=1,2,3,4,5$ respectively.

To analyze the dependence on the boundary parameter $\gamma$, we consider its asymptotic behavior. Using the expansion~\cite{gradshteyn-ryzhik}
\begin{equation}
    \Gamma(\alpha,y)\to y^{\alpha-1}e^{-y}+\mathcal{O}(1/y),\quad |y|\to\infty,
\end{equation}
we find that
\begin{equation}
\lim_{\gamma \to \infty} \mathcal{P}^{+}(\Omega) = 0,
\end{equation}
which reflects the expected suppression of unstable modes under Dirichlet boundary conditions.

\begin{figure}[!htb]
    \centering
    \includegraphics[width=0.9\linewidth]{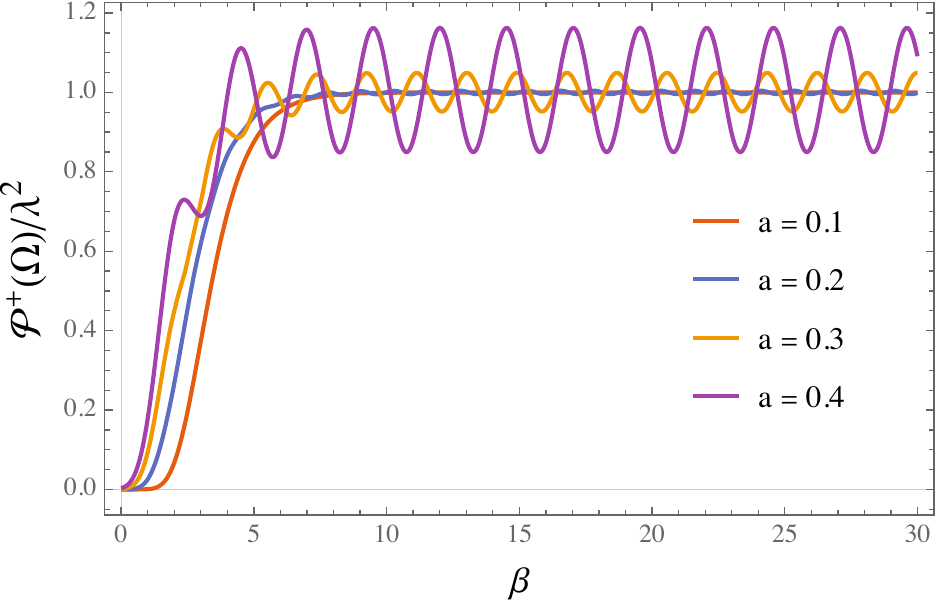}
    \caption{Transition probability $\mathcal{P}^{+}(\Omega)/\lambda^2$ given by Eq.~\eqref{probability4} as a function of $\beta = \log(1/ \gamma)$, illustrating the behavior near the Neumann boundary condition ($\gamma \to 0$). Here, we are setting $\Omega = 1$.}
    \label{fig:prob4}
\end{figure}

To study the opposite regime, we reparametrize the boundary parameter as $\gamma := e^{-\beta}$, so that $\beta \to \infty$ corresponds to the Neumann limit. Fig.~\ref{fig:prob4} shows the transition probability as a function of $\beta$. Interestingly, as we approach the Neumann boundary condition, the response becomes increasingly oscillatory.

The persistence of finite and oscillatory responses under acceleration suggests that non-inertial motion may act as an energy regulator for the non-oscillatory free particle modes permitted by the Neumann boundary condition. Since acceleration effectively injects energy into the system, it can modify the coupling between the detector and the arbitrarily low-frequency (non-oscillatory) mode, thereby avoiding the infrared divergence that arises in the inertial case.

\section{Conclusions}\label{sec:final-remarks}

In this work, we have investigated the emergence of a single unstable mode induced by a Robin boundary condition at a naked singularity in (1+1)-dimensional half-Minkowski spacetime. The unstable mode is controlled by a boundary parameter $\gamma$, which mixes Dirichlet ($\gamma \to \infty$) and Neumann ($\gamma \to 0$) boundary conditions. We quantized this unstable mode within the rigged Hilbert space (RHS) formalism, where the corresponding quantum state is understood as a generalized eigenstate with imaginary energy. This framework enables the selection of a time domain in which the unstable mode exists and naturally describes it as a decaying (or growing) particle, thus curing the temporal divergence typically associated with its exponential behavior.

To probe this quantization scheme, we analyzed how an observer would interact with such states using the Unruh–DeWitt detector model. By isolating the contribution from the unstable mode, we computed the detector’s transition probability for three representative trajectories: static, inertial, and uniformly accelerated.

For static observers, the detector response exhibits a resonance profile of the Breit–Wigner form, with the boundary parameter $\gamma$ directly controlling the decay width $\Gamma_0 = 2\gamma$, and hence the mean lifetime $\tau_{\text{mean}} \sim 1/\Gamma_0$. Moreover, the probability amplitude is exponentially suppressed with increasing distance from the boundary. In the Dirichlet limit ($\gamma \to \infty$), the detector remains unexcited, indicating the absence of a decaying particle. In contrast, the Neumann limit ($\gamma \to 0$) yields a free, non-oscillatory particle with arbitrarily low frequency, leading to an infrared-divergent detector response.

For inertial observers moving at constant velocity $v$, the qualitative dependence on $\gamma$ remains unchanged, but the decay width experiences a Doppler shift $\Gamma_v = \Gamma_0 \sqrt{(1+v)/(1-v)}$, making the mean lifetime velocity-dependent. In the ultrarelativistic limit $v \to 1$, the decay becomes instantaneous, and the detector registers no excitation.

In the accelerated case, the detector’s response takes a more intricate form governed by incomplete gamma functions. Although no simple Breit–Wigner structure emerges, the transition probability remains peaked around $\Omega=0$ and exhibits nontrivial dependence on the acceleration $a$. Notably, in the Neumann limit, the infrared divergence of the inertial motion is replaced by a finite, highly oscillatory response, suggesting that the energy injected by the detector’s acceleration acts as a physical regulator for the infrared pathology.

In summary, our results demonstrate that the combination of RHS quantization and detector-based probes offers a consistent and physically meaningful framework for describing unstable field configurations. This approach provides a natural interpretation of exponentially growing solutions as decaying quantum states and opens new avenues for investigating the interplay between instability and observers in the semiclassical theory. (

\acknowledgments

B. S. F. acknowledges support from the Conselho Nacional de Desenvolvimento Científico e Tecnológico (CNPq, Brazil), Grant No. 161493/2021-1. J. P. M. P. thanks the support provided in part by Conselho Nacional de Desenvolvimento Científico e Tecnológico (CNPq, Brazil), Grant No. 305194/2025-9.

\end{document}